\documentclass[preprint,amsmath,amssymb,pre,superscriptaddress]{revtex4-1}
\usepackage{color,longtable}
\usepackage{graphicx}
\usepackage{amssymb,amsfonts,amsmath}
\usepackage{here}

\begin{document}
\title{Effective osmotic pressure of foams in short-term dynamics}
\author{Aoi Kaneda}
\author{Rei Kurita}
\email{kurita@tmu.ac.jp}
\affiliation{%
Department of Physics, Tokyo Metropolitan University, 1-1 Minamioosawa, Hachiouji-shi, Tokyo 192-0397, Japan
}%
\date{\today}
\begin{abstract}
Foams exhibit absorptive properties and are widely used for cleaning contaminants, oil recovery, and selective mineral extraction via froth flotation. Although foam absorption has historically been linked to equilibrium osmotic pressure, empirical observations show that drainage occurs at levels much lower than theoretical predictions. In this study, we investigate the effective osmotic pressure in foams. The experimental findings indicate that effective osmotic pressure is primarily influenced by short-term dynamics and is governed by the yield stress rather than equilibrium osmotic pressure. Gravity-induced flow drives bubble movement downward, subjecting the lower bubbles to increased pressure. Once the pressure exerted by the solution exceeds the yield stress, internal bubble rearrangement occurs, and drainage is promoted by the relaxation of inter-bubble repulsive forces. In contrast, below the yield stress, these repulsive forces suppress drainage. While foams have traditionally been modeled as immobile porous media, the reality involves kinematic coupling between the solution and the bubbles, which establishes yield stress as the key factor in determining effective osmotic pressure. This framework is also relevant to the dynamics of soft jammed systems, such as blood flow in vessels, emulsions, and biological tissues, offering significant advancements in the understanding of soft jammed system behavior.
\end{abstract}


\maketitle

\section{Introduction}
Foam is composed of tightly packed air bubbles within a liquid, creating a functional state that combines gas-like properties for insulating mass and heat transport, liquid-like properties for absorbing water and oil, and solid-like properties for maintaining shape. Due to this versatility, foams are widely used across industries, including in cleaning agents, food products, fire extinguishing systems, oil recovery, and selective mineral extraction through froth flotation~\cite{Weaire2001, Cantat2013}. A key physical property of foams is osmotic pressure~\cite{Weaire2001, Cantat2013, Princen1986, Princen1987, Mason1997, Hohler2008, Yanagisawa2021a, Yanagisawa2023}. According to Laplace's equation, a pressure difference $\Delta P$ arises between the inside and outside of a surface with curvature:  $\Delta P = \gamma (1/r_1 + 1/r_2)$, where $\gamma$ is the surface tension, and $r_1$ and $r_2$ are the principal radii of curvature. In foams, the plateau borders have negative curvature, resulting in negative pressure, whereas the liquid film has a flat curvature. This pressure differential between the liquid film and the plateau borders generates the overall negative osmotic pressure of the foam, allowing it to absorb water, oil, and microorganisms~\cite{Keyvan2013, Roveillo2020}.

The osmotic pressure of emulsions, which are analogous to foam systems, was first measured in pioneering experiments by Princen~\cite{Princen1986, Princen1987}. In these studies, a semipermeable membrane was placed beneath the emulsion, and the osmotic pressure $\pi$ was calculated by balancing the weight of the liquid with the osmotic pressure~\cite{Princen1987}. The vertical osmotic pressure gradient at a height $z$ is given by:
\begin{eqnarray}
[\nabla \pi (z)]_z = \rho g \phi (z), 
\end{eqnarray}
where $[\nabla \pi (z)]_z$ represents the osmotic pressure gradient in the vertical direction, and $\rho$, $g$, and $\phi (z)$ are the solution density, gravitational acceleration, and the liquid fraction of the foam at height $z$, respectively. Integrating this expression from 0 to $H$, the following equation is obtained:
\begin{eqnarray}
\pi(H) - \pi(0) = \int^H_0 \rho g \phi (z) dz.\label{eq:int}
\end{eqnarray}
Here, $H$ is the height from the base of the foam, and $\pi (0) = 0$ since the foam reaches its jamming point at the bottom. In monodisperse systems, the dependence of osmotic pressure $\pi$ on the liquid fraction $\phi$ has been established, showing that osmotic pressure scales with the mean bubble radius $R$ and the liquid fraction $\phi$~\cite{Weaire2001, Cantat2013}:
\begin{eqnarray}
\pi = \frac{\gamma}{R}\sqrt{\frac{0.33}{\phi}} \label{eq:phi}
\end{eqnarray}
In systems where bubble sizes vary, this expression slightly diverges, although the order of magnitude for the osmotic pressure remains similar~\cite{Mason1997, Hohler2008}.

In reality, as the liquid fraction decreases due to drainage, the liquid film in the foam collapses. The maximum osmotic pressure, $\pi_{max}$, is determined by the uppermost region of the foam. Let the initial liquid fraction be $\phi_0$ and the maximum height be $H_{max}$. Assuming minimal evaporation and conserving the mass of water, the following condition holds:
\begin{eqnarray}
\frac{1}{H_{max}} \int^{H_{max}}_0 \phi(z) dz = \phi_0.
\end{eqnarray}
By combining this equation with Eq.~\ref{eq:int}, we arrive at:
\begin{eqnarray}
\pi (H_{max}) = \rho g H_{max} \phi_0 \label{eq.Pi}
\end{eqnarray}
This equation indicates that the osmotic pressure at the top of the foam determines the maximum amount of liquid that can be retained per unit cross-sectional area of the foam. For instance, if the surface tension is 37 mN/m, the mean diameter of the bubbles is 0.3 mm, and the liquid fraction at the collapse threshold of the liquid film is 0.5\%, then $\pi_{max}$ is approximately 2000 Pa. Using Eq.\ref{eq.Pi}, this can be converted into a foam height with an initial liquid fraction of 10\%, which is around 1 meter. If the foam height is below this value, the solution will not drain out. However, in practice, solution drainage occurs in much smaller foams~\cite{Tani2022}, revealing a significant discrepancy between theoretical predictions and actual behavior. The underlying reasons for this gap remain largely unresolved, and there has been little in-depth research or discussion on the factors that govern the drainage limit.

In this study, the effective osmotic pressure $\pi_{eff}$, which differs significantly from $\pi_{max}$, was quantitatively measured using three surfactants, with the initial liquid fraction and bubble size as parameters. It was observed that $\pi_{eff}$ is over two orders of magnitude lower than the theoretical value derived from $\pi_{max}$. During drainage, internal bubbles were seen to rearrange, while they remained stationary in the absence of drainage, indicating a relationship between drainage and yield stress. Moreover, $\pi_{eff}$ remained nearly constant even when foams became more fragile due to salt addition. This observation reinforces the notion that the drainage limit is governed by yield stress rather than maximum osmotic pressure or minimum liquid film thickness. Our findings suggest that the deformation and rearrangement of internal bubbles in conjunction with the motion of the solution create a non-equilibrium state that is significantly different from the equilibrium state, leading to a marked decrease in the drainage limit of the foam. Since these characteristics are likely prevalent in jammed systems with soft particles, our results offer valuable insights into the dynamic properties of condensed emulsions, proteins, and biological tissues.

\section{Materials and Methods}
\subsection{Foam}
We utilized a 5.0 wt\% solution of the ionic surfactant TTAB (tetradecyltrimethylammonium bromide) mixed with glycerol and deionized water. This concentration exceeds the critical micelle concentration of 0.12 wt\%~\cite{Danov2014}, and the surface tension at this concentration is measured at 37 mN/m. It is established that the interface rigidity of TTAB is minimal~\cite{Yanagisawa2023}. To prevent foam collapse, a 20 wt\% glycerol solution was employed. The density of this solution, $\rho$, is 1.10 g/cm$^3$, while the viscosity, $\eta$, of the 20 wt\% glycerol solution is 2.7 mPa$\cdot$s.
Additionally, we employed a 5.0 wt\% solution of the ionic surfactant SDS (sodium dodecyl sulfate) in glycerol and deionized water. The surface tension for this solution is 39.7 mN/m, with $\rho$ = 1.08 g/cm$^3$, and $\eta$ of the 30 wt\% glycerol solution is 2.8 mPa$\cdot$s~\cite{Castro1998}.
We also tested a household detergent (Charmy, Lion Co., Japan) diluted to 20 wt\% with deionized water. This detergent comprises several surfactants, including sodium tetradecenesulfonate, polyoxyethylene fatty acid alkanolamide, alkylamine oxide, sodium alkyl ether sulfate, and polyoxyethylene alkyl ether. The density of this solution is 1.00 g/cm$^3$, with a surface tension of 25 mN/m and a viscosity of 1.68 mPa$\cdot$s.

\begin{figure}[htbp]
\centering
 \includegraphics[width=8cm]{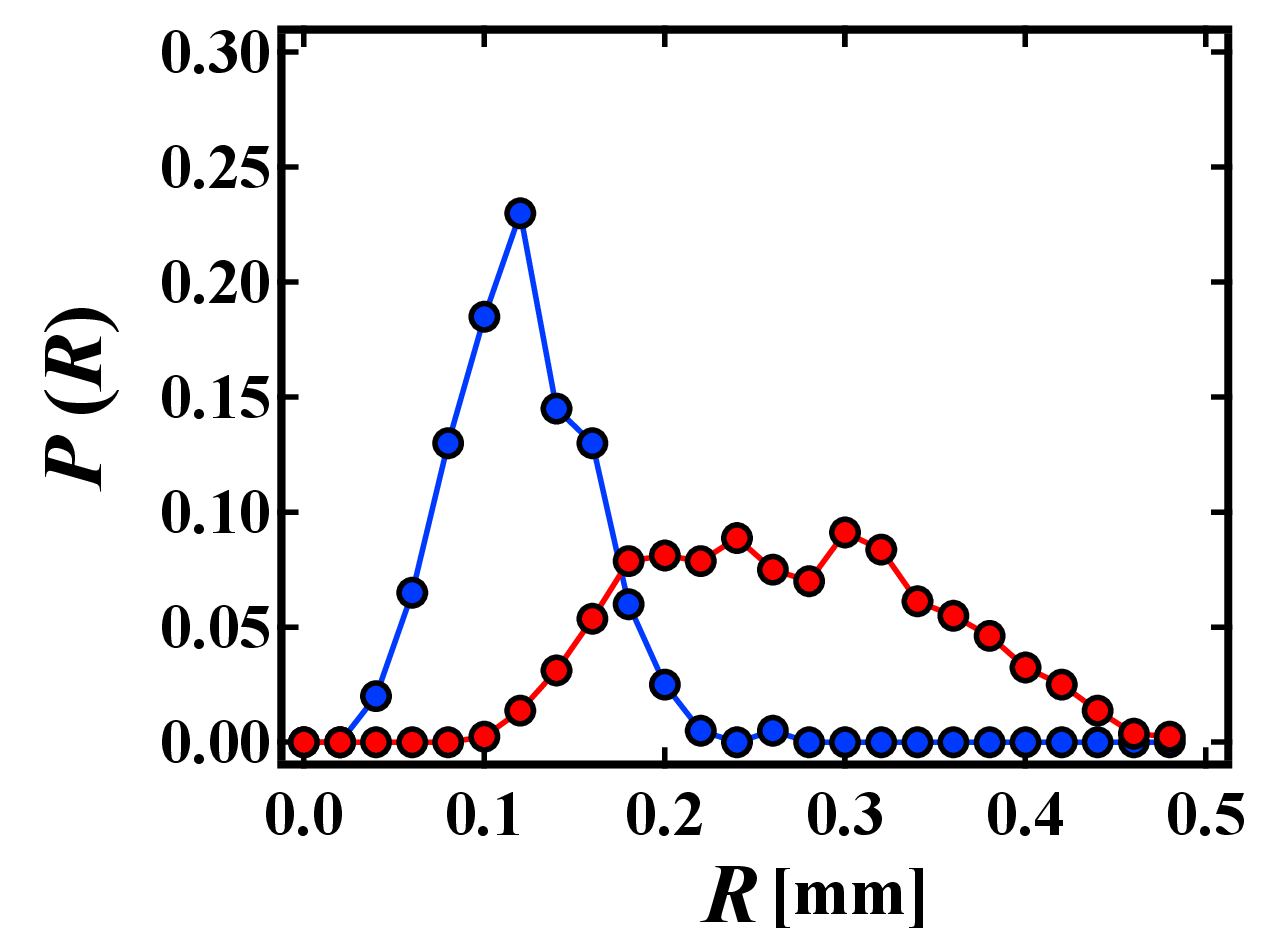}
\caption{The size distributions of bubbles in foams prepared using two different dispensers were measured 5 minutes after pumping out, with Charmy surfactant utilized. The mean bubble radius $R$ for the foam with smaller bubbles was 0.131 mm, accompanied by a standard deviation of 0.037 mm. In contrast, the mean bubble radius $R$ for the foam with larger bubbles was found to be 0.283 mm, with a standard deviation of 0.082 mm. }
\label{distribution}
\end{figure}

\begin{figure}[htbp]
\centering
\includegraphics[width=8cm]{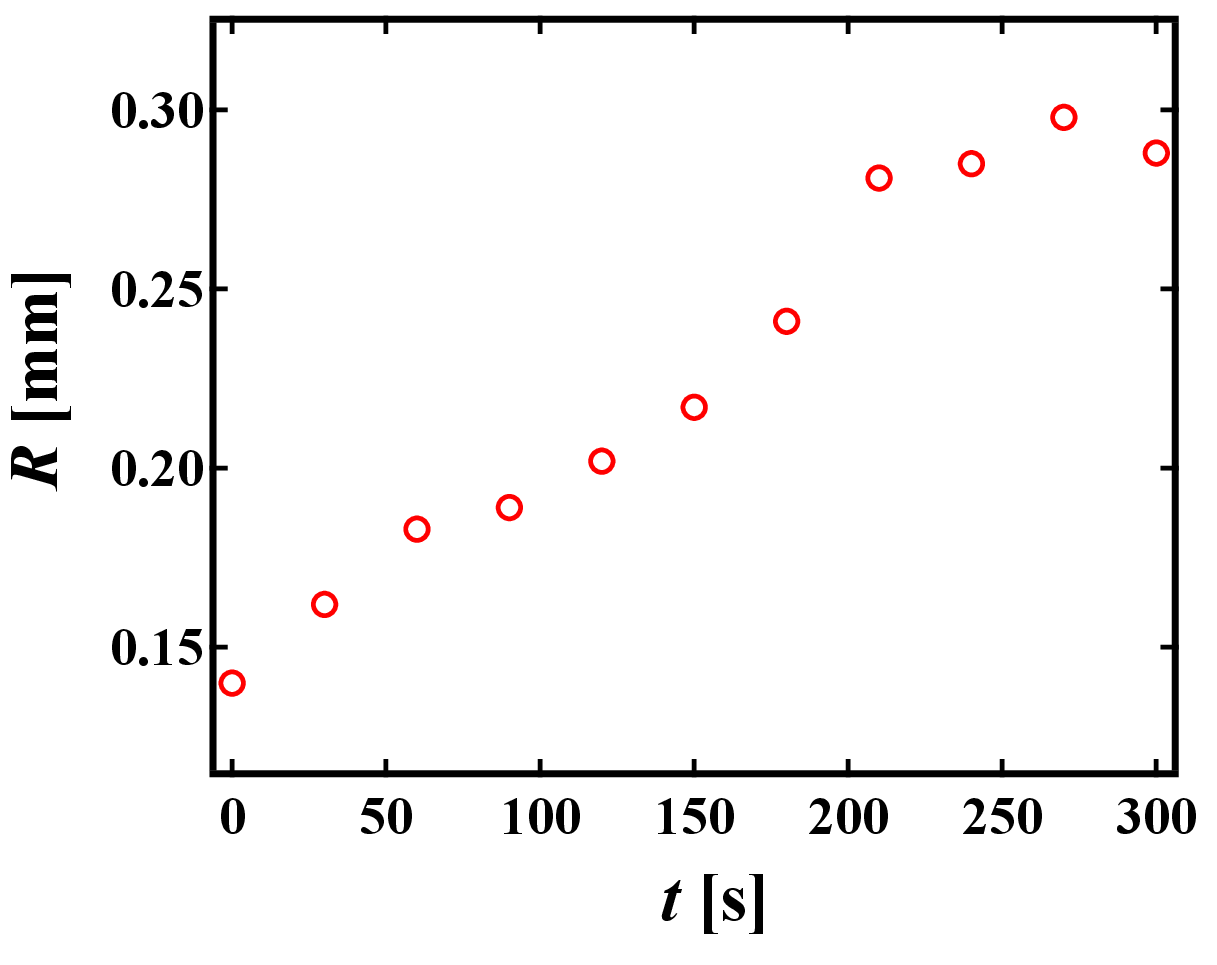}
\caption{The time evolution of the mean bubble radius $R$ due to coarsening is shown for foams with Charmy surfactant. The bubble size steadily increases within the first 5 minutes after foam generation.}
\label{coarsening}
\end{figure}

Two distinct types of foam dispensers were employed to generate foams with varying bubble sizes. Smaller bubbles were produced using a foam dispenser from Awahour (Torigoe Plastic Ind. Co. Ltd., Japan), while larger bubbles were generated with a dispenser from Daiso Ind. Co., Ltd. (Japan). Figure \ref{distribution} depicts the bubble size distributions for the two types of foam after 5 minutes of generation.
The mean bubble radius $R$ for the smaller bubble foam was determined to be 0.131 mm, with a standard deviation of 0.037 mm. In contrast, the mean radius for the larger bubble foam was 0.283 mm, with a standard deviation of 0.082 mm. Bubble sizes were measured through image analysis for each sample. Although previous studies frequently utilized insoluble gases (CDC) to mitigate coarsening, our approach involved using air bubbles, reflecting more realistic conditions. Consequently, the bubble size increases over time due to coarsening, as illustrated in Fig.~\ref{coarsening}.

The mean three-dimensional liquid fraction $\phi$ was calculated using the equation $\phi = m/{\rho}V_{foam}$, where $m$ is the mass of the liquid, $\rho$ is the liquid density, and $V_{foam}$ is the foam volume. The measurement error for the liquid fraction is 0.005.

\subsection{Experimental setup}
\begin{figure}[htbp]
\centering
\includegraphics[width=8cm]{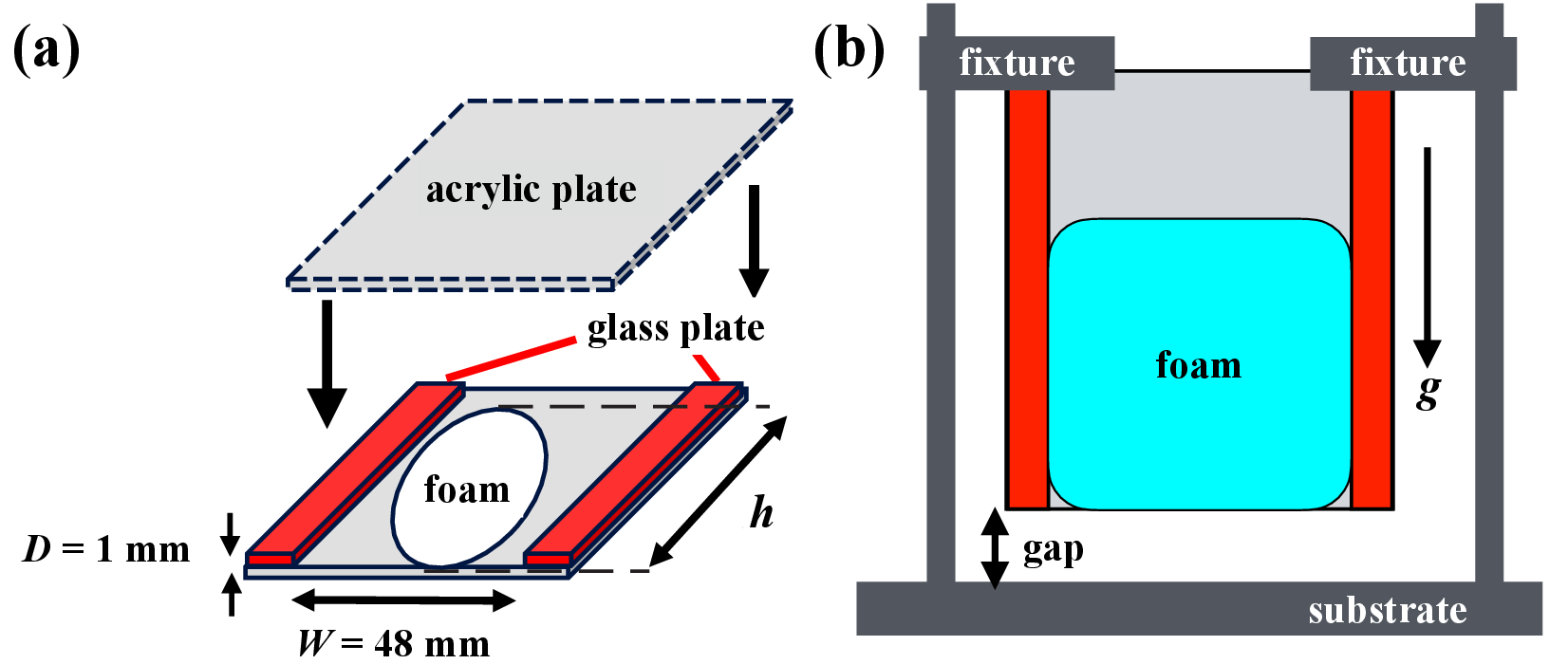}
\caption{Schematic of the experimental setup. (a) Foam is confined between two large acrylic plates. The gap of the Hele-Shaw cell, $D$, was set to either 1 mm or 2 mm using spacers attached to the acrylic plates. The inner width, $W$, is 48 mm. (b) The foam was introduced into the Hele-Shaw cell, which was then positioned vertically. A gap was created between the cell and the substrate to prevent water leakage caused by capillary action.
}
\label{setup}
\end{figure}

A schematic of the experimental setup is provided in Fig.~\ref{setup}. The Hele-Shaw cell was constructed using two acrylic plates, with the gap between the plates $D$ set to either 1 mm or 2 mm, maintained by spacers affixed to the plates. The internal width of the compartment was $W$ = 48 mm, while the foam height $H$ and the mean liquid fraction $\phi_0$ were adjustable parameters. The foam was placed inside the Hele-Shaw cell, which was then positioned vertically. A gap was introduced between the cell and the substrate to prevent water leakage via capillary action. Foam behavior was recorded at the center of the cell, where the spacer influence was minimal, and images were captured using a digital camera (EOS R, Canon Inc., Japan).

\section{Results}
\subsection{Diagram}
\begin{figure}[htbp]
\centering
\includegraphics[width=8cm]{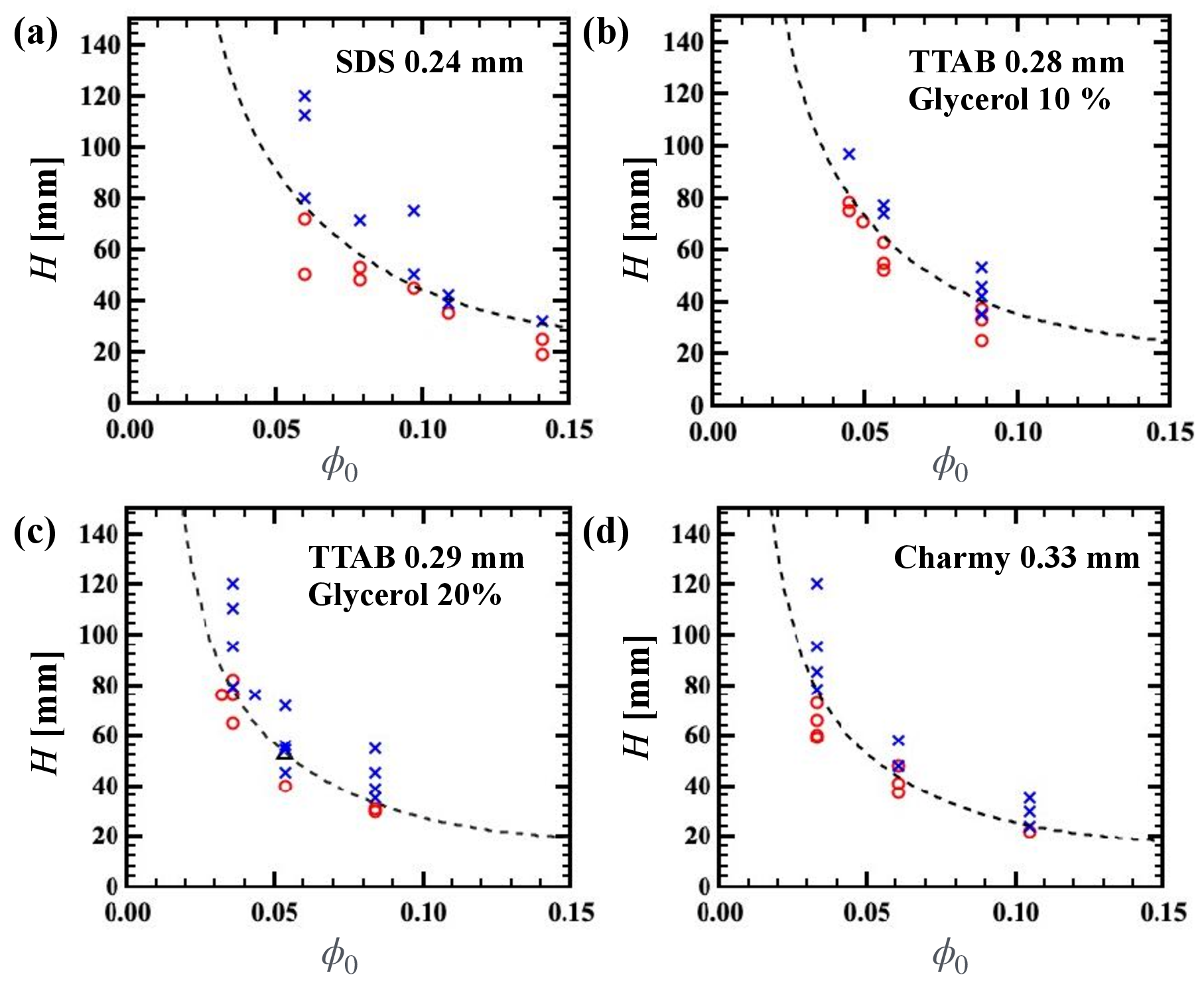}
\caption{Diagrams showing drainage as a function of $\phi_0$ and $H$. The surfactants used and the corresponding mean bubble diameters are (a) SDS, 0.18 mm, (b) TTAB with 10\% glycerol, 0.28 mm, (c) TTAB with 20\% glycerol, 0.29 mm, and (d) Charmy, 0.33 mm. Circle symbols indicate undrained conditions, while cross symbols indicate drained conditions. The dotted line in each diagram represents the threshold separating drained and undrained regions, which can be described by the relationship $H\phi_0$ = constant. This boundary correlates with the maximum effective osmotic pressure $\pi_{eff}$.
}
\label{diagram}
\end{figure}

We first examined the foam's drainage behavior as a function of its initial liquid fraction $\phi_0$ and height $H$. Drainage was considered to have occurred if, after 20 minutes, a 1 mm thick layer of solution accumulated beneath the lowest bubble; if not, the foam was classified as undrained. Figure~\ref{diagram} presents the drainage state diagram. The surfactants and corresponding mean bubble sizes were as follows: (a) SDS, 0.18 mm, (b) TTAB with 10\% glycerol, 0.28 mm, (c) TTAB with 20\% glycerol, 0.29 mm, and (d) Charmy, 0.33 mm. Circles indicate undrained conditions, while crosses indicate drained conditions. The dotted line in the diagram marks the boundary between drainage and non-drainage, which can be expressed as $H\phi_0$ = constant. The drainage behavior near this boundary was analyzed in detail.

\begin{figure}[htbp]
\centering
 \includegraphics[width=8cm]{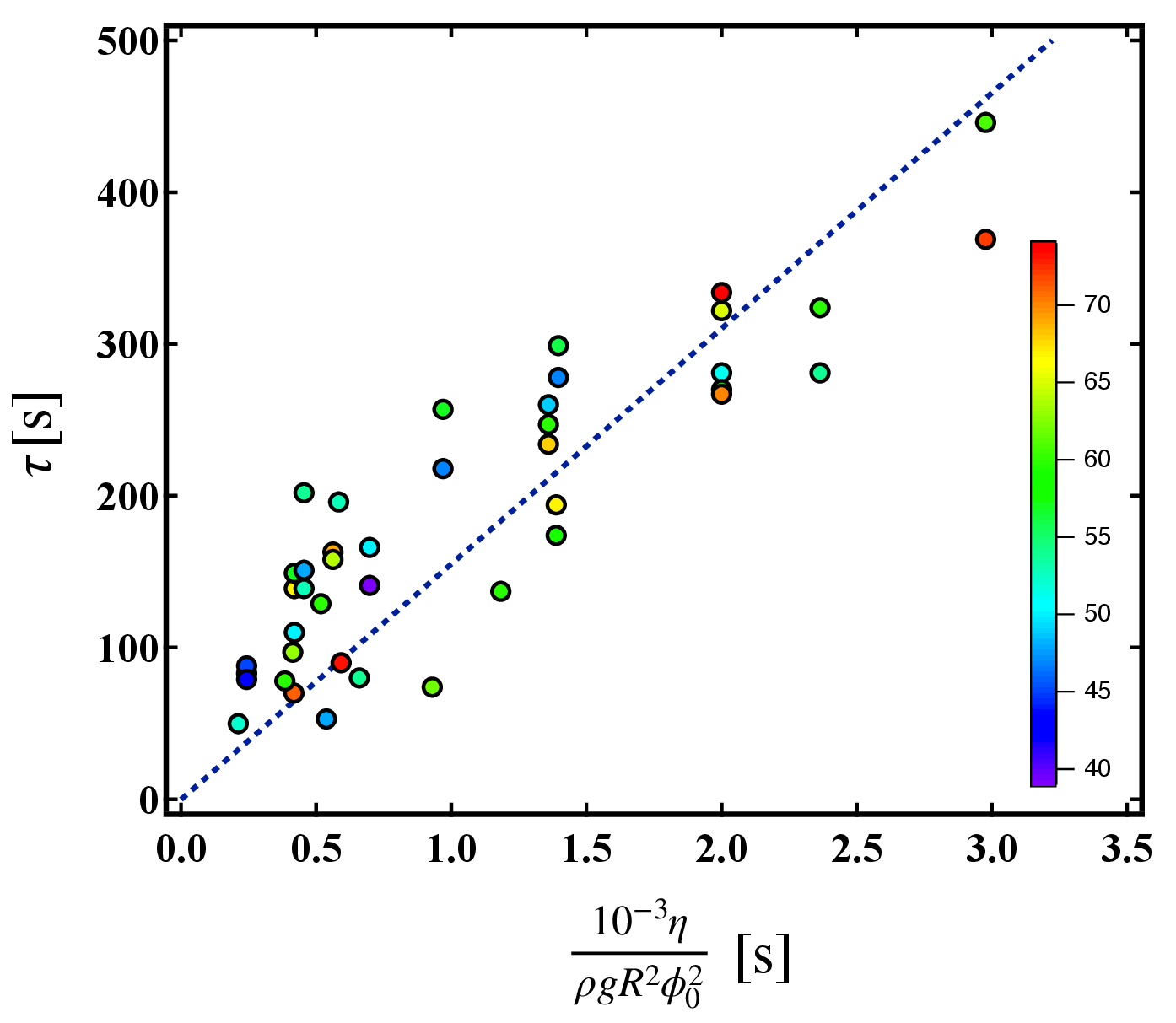}
\caption{
The correlation between $\tau$ and $\eta/R^2 \rho g \phi_0^2$ for TTAB foam under drained conditions demonstrates a proportional relationship, consistent with theoretical predictions. The color in the plot represents $ \tau $ for varying foam heights $H$. There is little correlation between $ \tau $ and $H$, which aligns with expectations since the osmotic pressure gradient is not significant during the early stages of drainage~\cite{Cantat2013}.
}
\label{drainage}
\end{figure}

Next, we investigated the drainage time within the drainage regime. The drainage time at the earlier stage is theoretically described using Darcy's law for flow through pores~\cite{Cantat2013, Koehler1998, deGennes2013}. The drainage front velocity $u$ in foam can be expressed as:
\begin{eqnarray}
u = C_c \frac{\rho g R^2 \phi_0}{\eta}, \label{eq:u}
\end{eqnarray}
where $C_c$ is a constant dependent on the Boussinesq number $B_o$~\cite{Cantat2013, Leonard1965}. The Boussinesq number $B_o$ quantifies the ratio of surface viscosity to bulk viscosity and is defined as $B_o = \eta_s/\eta r$, where $r$ is the radius of the Plateau border. In our experiment, the drainage time $ \tau $ was measured when the solution reached a height of 1 mm, which corresponds to $1/\phi_0$ mm inside the foam. Thus, the velocity $u$ was determined as $u = 10^{-3}/\phi_0\tau$ m/s.
Substituting into Eq.~\ref{eq:u}, we obtain:
\begin{eqnarray}
\tau = \frac{10^{-3} \eta}{C_c R^2 \rho g \phi_0^2},
\end{eqnarray}
Figure~\ref{drainage} shows the relationship between $\tau$ and $\eta/R^2\rho g \phi_0^2$, confirming a proportional correlation consistent with theoretical predictions. The slope gives $C_c = 0.00735$, which corresponds to $B_0 \sim 2$ according to previous studies~\cite{Cantat2013}. Using TTAB's surface viscosity $ \eta_s = 5 \times 10^{-8}$ kg/s~\cite{Lorenceau2009}, bulk viscosity $ \eta = 2.7$ mPa$\cdot$s, and a Plateau border radius $r = 20 \mu$m, we calculate $B_0 \sim 0.9$, which is in close agreement.
We also examined $\tau$ at varying foam heights $H$, and color coding was used to indicate $\tau$ for different $H$. There was little correlation between $\tau$ and $H$. Notably, in our experiment, $\tau$ was at most 7 minutes, significantly shorter than the measurement time for equilibrium osmotic pressure (several days)~\cite{Princen1987}. This is expected since the osmotic pressure gradient is irrelevant during the early stages of drainage.
Above the boundary line, drainage did not occur even after 30 minutes, suggesting a sharp rather than a gradual boundary between drained and undrained states.

\begin{figure}[htbp]
\centering
 \includegraphics[width=8cm]{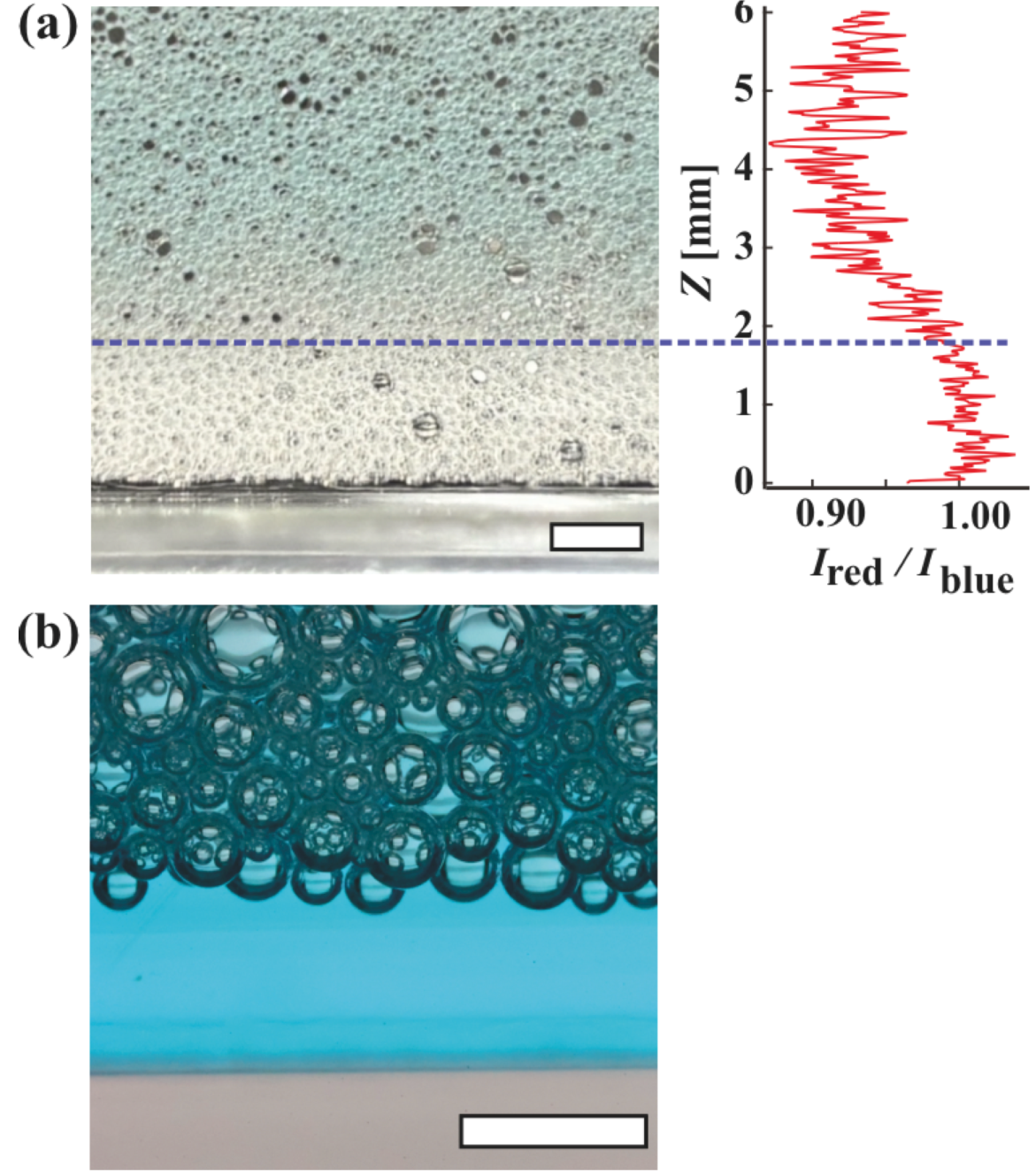}
\caption{
An image taken five minutes after the solution made contact with the TTAB foam, where the mean bubble radius $R = 0.20$ mm. The foam height $H$ is 38 mm, and the initial liquid fraction $\phi_0$ is 0.03. To enhance the visibility of the solution penetration, the solution within the foam was dyed blue.
For a quantitative analysis of the absorption distance, the image was split into RGB channels. The intensity of red, $I_{red}$, the complementary color to blue, was normalized by the blue intensity $I_{blue}$. The ratio $I_{red}/I_{blue}$ indicates that the solution was absorbed up to a depth of approximately 1.8 mm.
}
\label{image}
\end{figure}

Meanwhile, we examined the foam behavior below the boundary line in Fig.\ref{diagram} when a small amount of solution came into contact with the foam from below. Figure\ref{image} displays an image taken five minutes after the solution interacted with the foam. To visually track the penetration, the solution was colored blue.
For a quantitative analysis of the absorption distance, the color image was separated into RGB channels, and the red intensity $I_{red}$, the complementary color to blue, was normalized by the blue intensity $I_{blue}$. The reduction in $I_{red}/I_{blue}$ starting around $z = 1.8$ mm indicates that the solution was absorbed up to approximately 1.8 mm.
It is worth noting that the maximum height of solution rise due to capillary action, $H_c$, is about 0.15 mm, calculated using the equation $H_c = 2 \gamma \cos \theta / W \rho g$, where $ \theta $ represents the wetting angle. Since the absorption distance in the foam is significantly larger than the capillary rise in the Hele-Shaw cell, the solution's rise is not driven by capillary action but by absorption into the foam.

From these results, it is evident that the boundary line $H \phi_0 = \text{const.}$ reflects the balance between absorption and drainage. As this phenomenon closely resembles the measurement of static osmotic pressure~\cite{Princen1987}, we term it the effective osmotic pressure $\pi_{eff}$. Using Eq.~\ref{eq.Pi}, the maximum value of $\pi_{eff}$ is estimated to be at most 70 Pa from the boundary.
In contrast, the maximum osmotic pressure $\pi_{max}$ can be approximated at 2000 Pa, assuming a surface tension of 37 mN/m, an average bubble diameter of 0.3 mm, and a liquid fraction of 0.5\% at the film rupture limit. The significant discrepancy, about 30 times, suggests that the effective osmotic pressure differs substantially from the static osmotic pressure.

\subsection{Bubble size dependence of the effective osmotic pressure}
\begin{figure}[htbp]
\centering
 \includegraphics[width=8cm]{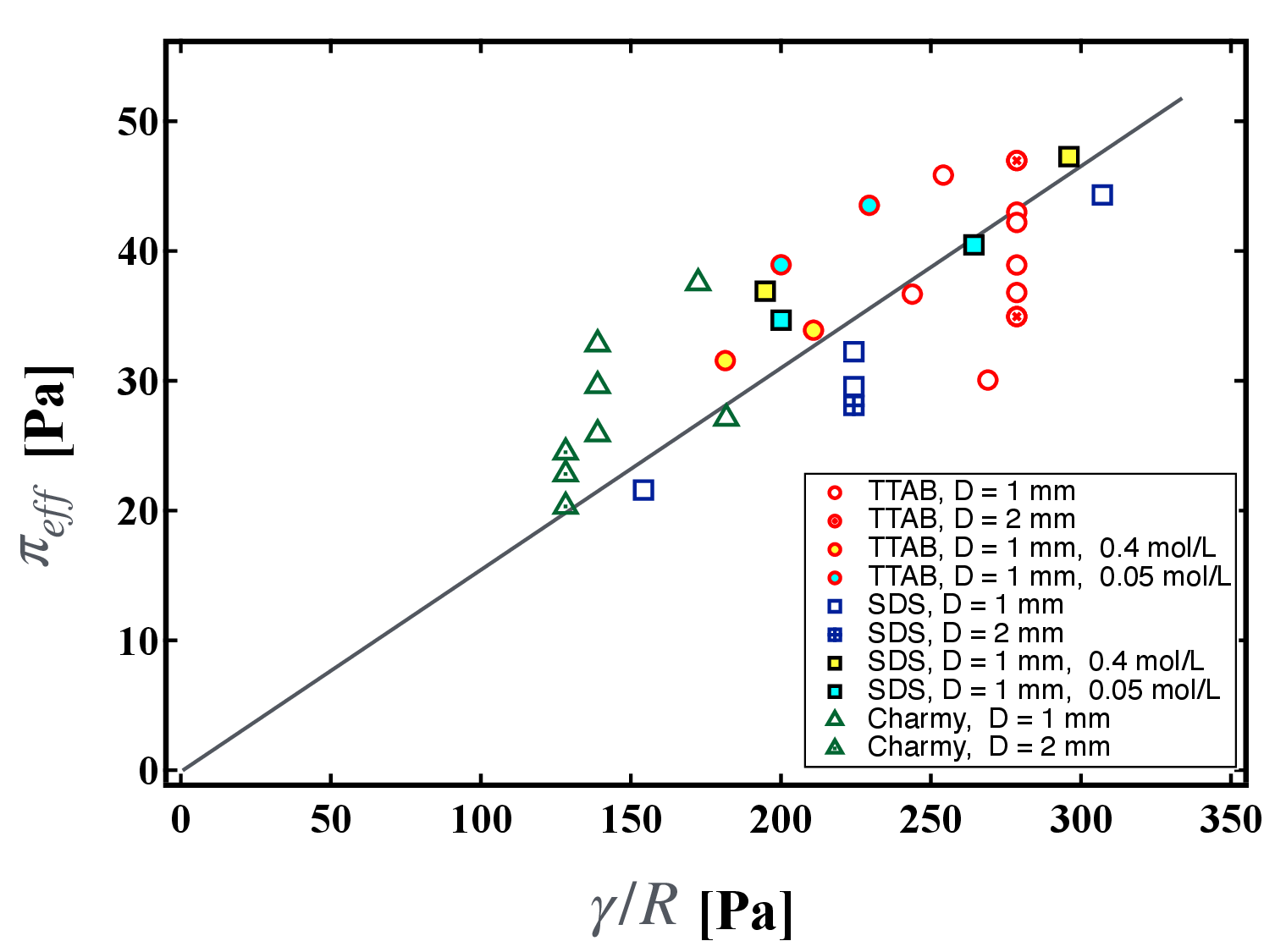}
\caption{The dependence of $\pi_{eff}$ on $\gamma/R$, where the circle, square, and triangle symbols correspond to $\pi_{eff}$ for TTAB, SDS, and Charmy, respectively. The symbols with midpoints represent $\pi_{eff}$ when the Hele-Shaw cell thickness is increased to $D$ = 2 mm, and these results are consistent with those obtained for $D$ = 1 mm. The filled symbols represent $\pi_{eff}$ after the addition of salt, showing that $\pi_{eff}$ has little correlation with the equilibrium osmotic pressure. The straight line is an approximation given by $\pi_{eff} = 0.156 \gamma/R$.
}
\label{scaling}
\end{figure}

Next, we explore the bubble size dependence of $\pi_{eff}$ to gain insight into its characteristics. Figure \ref{scaling} presents the relationship between $\pi_{eff}$ and $\gamma/R$. The circle, square, and triangle markers correspond to $\pi_{eff}$ measurements for TTAB, SDS, and Charmy, respectively. Glycerol concentrations of 5 \%, 10 \%, and 20 \% were used, and it was observed that $\pi_{eff}$ remains independent of viscosity.
The Pearson correlation coefficient between $\pi_{eff}$ and $\gamma/R$ is 0.774, indicating a strong correlation, despite some variation among the samples. The relationship was found to be approximately linear, scaled by $\gamma/R$ across different surfactants, and can be described by $\pi_{eff} = 0.156 \gamma/R$.
Additionally, the midpoint symbol represents $\pi_{eff}$ when the thickness of the Hele-Shaw cell is altered to $D = 2$ mm, which is consistent with the results for $D = 1$ mm. This suggests that the effect of the Hele-Shaw cell thickness is negligible.

We also measured $\pi_{eff}$ after adding NaCl to the surfactant solution. Since TTAB is a cationic surfactant and SDS is an anionic surfactant, electrostatic repulsion between the liquid films occurs. The addition of salt leads to electrostatic shielding, which causes the liquid film to thin~\cite{Behera2014, Jiang2020}.
In our system, the addition of salt accelerates bubble collapse, resulting in an increase in $\phi_{min}$. Consequently, $\pi_{max}$ decreases according to Eq.\ref{eq:phi}. However, as shown by the filled symbols in Fig.~\ref{scaling}, $\pi_{eff}$ remains unchanged after salt is added. This confirms that $\pi_{eff}$ has little correlation with the equilibrium osmotic pressure.

\subsection{Structural rearrangement}
\begin{figure}[htbp]
\centering
 \includegraphics[width=8cm]{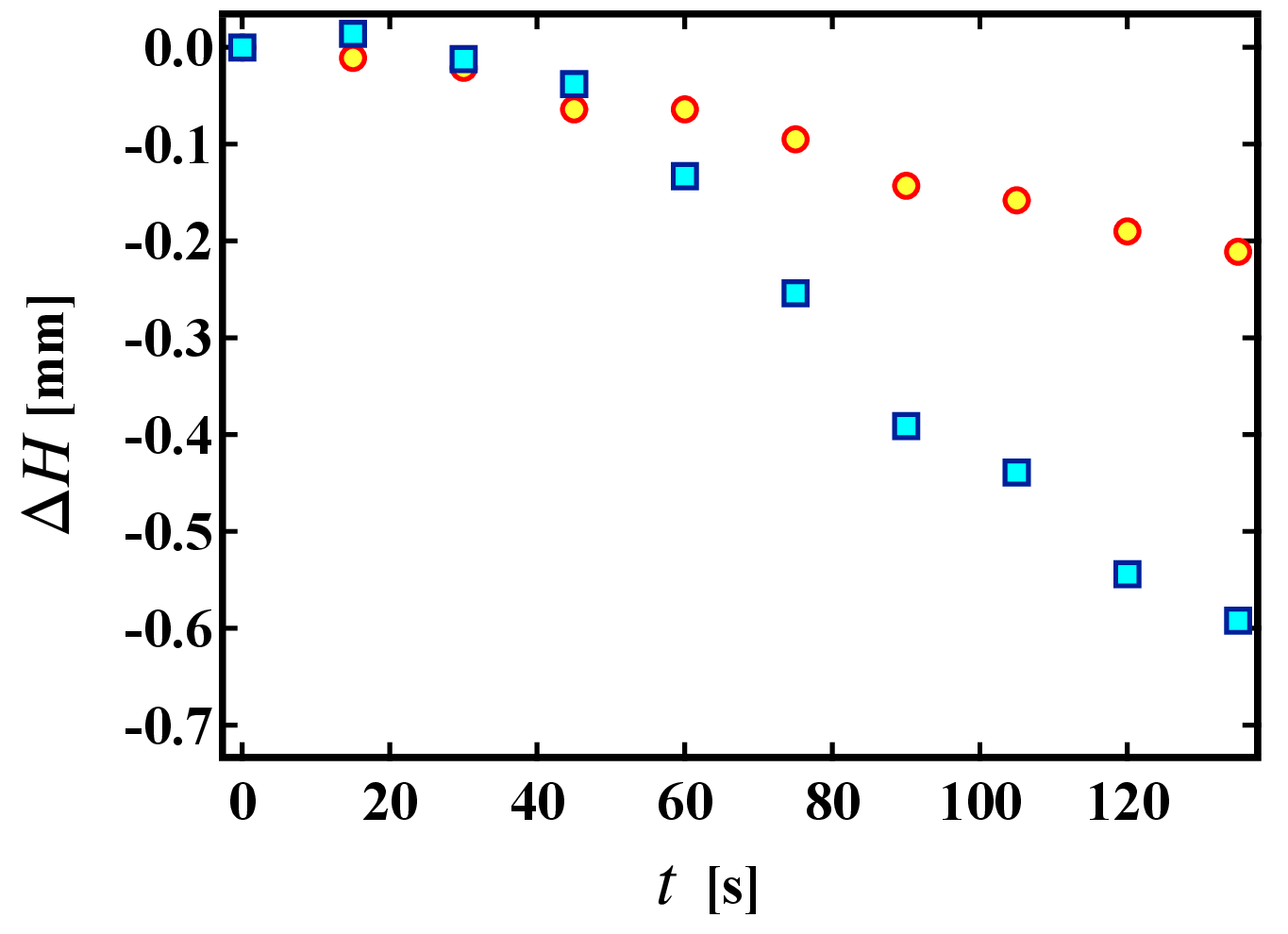}
\caption{The time evolution of foam height with drainage (circle) and without drainage (triangle). For non-draining conditions with TTAB, where the mean bubble diameter is 0.237 mm, $\phi_0$ = 9.6 \%, and $H$ = 37 mm, the foam height decreases by only 0.20 mm after 120 seconds. In contrast, under drained conditions with $H$ = 42 mm, the foam height decreases by 0.54 mm over the same period. The height reduction in the drained case is 2.7 times greater compared to the non-drained case.
}
\label{height}
\end{figure}

To explore the origin of the effective osmotic pressure, we examined the internal structures of foams both with and without drainage. After filling the bottom of the acrylic plate with foam and placing it vertically, we observed the foam height over time, as shown in Fig.~\ref{height}.
For non-draining conditions with TTAB, where the mean bubble diameter is 0.237 mm, $\phi_0$ = 9.6 \%, and $H$ = 37 mm, the foam height decreased by only 0.20 mm after 120 seconds. In contrast, under drained conditions with $H$ = 42 mm, the foam height dropped by 0.54 mm after the same period, indicating a 2.7 times greater reduction compared to the non-draining case. This significant change in strain with a minor change in solution mass suggests a transition-like phenomenon.

\begin{figure}[htbp]
\centering
 \includegraphics[width=8cm]{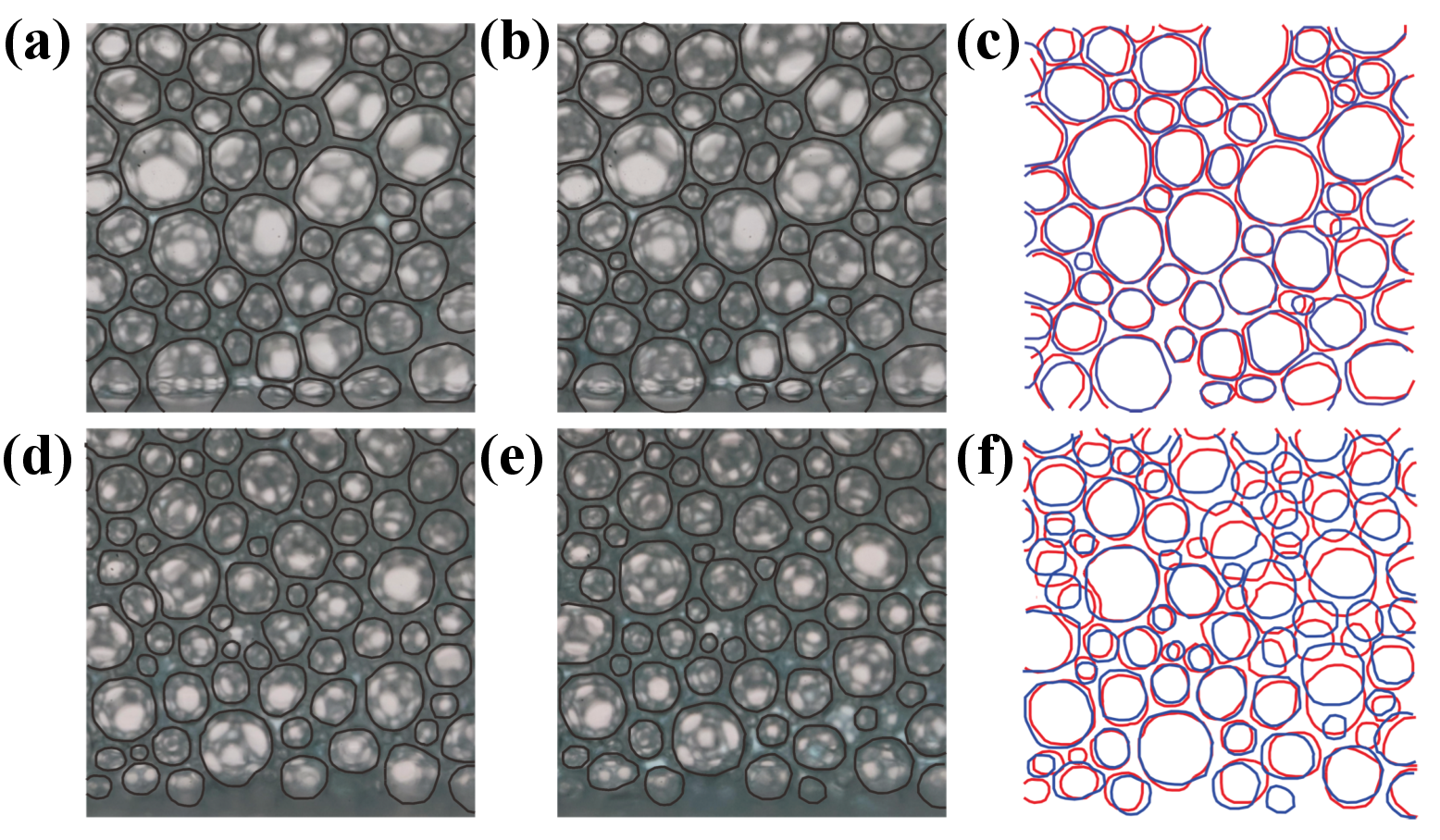}
\caption{The internal structural changes at the base of the foams were examined under both drained and non-drained conditions. Panels (a) and (b) illustrate the bubble positions at $t$ = 14.6 s and $t$ = 20.0 s for TTAB, with a mean radius of 0.18 mm and $H$ = 37 mm, indicating no drainage. Panel (c) presents the extracted and overlaid bubble interfaces from (a) and (b), showing that the bubbles remain stationary under the non-drained condition. In contrast, panels (d) and (e) depict the bubble positions at $t$ = 5.8 s and $t$ = 11.4 s for TTAB with $H$ = 42 mm under the drained conditions. Panel (f) shows the extracted and overlaid bubble interfaces from (d) and (e), revealing significant rearrangement of the bubbles during this time.
}
\label{rearrangement}
\end{figure}

Additionally, we examined the internal structural changes at the base of the foams under both drained and non-drained conditions. Figures \ref{rearrangement}(a) and (b) display the positions of the bubbles at $t$ = 14.6 s and $t$ = 20.0 s for TTAB, with a mean radius of 0.18 mm and $H$ = 37 mm, showing no drainage. In Figure \ref{rearrangement}(c), the extracted and overlaid bubble interfaces indicate that there is little rearrangement of the bubbles under non-draining conditions. Conversely, Figs.~\ref{rearrangement}(d) and (e) illustrate the bubble positions at $t$ = 5.8 s and $t$ = 11.4 s for TTAB with $H$ = 42 mm. During this period, significant rearrangement of the bubbles is observed, as depicted in Figure \ref{rearrangement}(f). This bubble rearrangement phenomenon occurs when the applied stress exceeds the yield stress. Consequently, $\pi_{eff}$ is expected to be related to the yield stress. In the wet domain, the yield stress is primarily influenced by $\gamma/R$ rather than $\phi$~\cite{Weaire2001,Weaire2008}. As shown in Fig.~\ref{scaling}, $\pi_{eff}$ can be scaled by $\gamma/R$, aligning with the scaling of the yield stress. The yield stress value is also known to be approximately 50 Pa, which is consistent with the findings of this experiment~\cite{Weaire2001,Weaire2008}. Moreover, it is noteworthy that the equilibrium osmotic pressure can be considered negligible in the early stages of drainage, as suggested by the scaling of the drainage time. These findings imply that the occurrence of drainage is determined by whether bubble rearrangement takes place.

\section{Discussion}
In this section, we explore the connection between yield stress and the drainage phenomenon. When the foam is positioned vertically, gravity causes the solution to flow downward. This flow generates a drag force on the bubbles, which are also pulled downward. Consequently, the pressure within the bubbles at the bottom rises, and this increase in pressure is believed to counteract drainage through a repulsive force. However, when the pressure exceeds the yield stress, the rearrangement of the bubbles occurs, leading to a reduction in pressure and a weakening of the repulsive force, thereby facilitating the drainage of the solution.

Next, we examine drainage in relation to time scales. In the short time scale, we have demonstrated that the establishment of a liquid fraction gradient in the $z$ direction is insufficient; rather, the pressure inhomogeneity arising from the kinematic coupling between the solution and the bubbles plays a more significant role. Conversely, in the long-term scenario, once the flow of the solution ceases, the stresses exerted on the bubbles by the flow dissipate, allowing structural relaxation to occur and thus resolving the pressure distribution. During this phase, instead of absorbing the drained liquid, the bubbles tend to migrate toward the drained solution, a phenomenon frequently observed in horizontal absorption experiments~\cite{Yanagisawa2021a, Yanagisawa2023}. 
Foams are characterized as jammed systems of bubbles, and such densely packed systems exhibit considerably long relaxation times, sometimes extending over several days. The osmotic pressure attained after this prolonged relaxation represents the equilibrium osmotic pressure, as identified by Princen et al~\cite{Princen1986}. Previous theories have overlooked the significance of $\pi_{eff}$ on short time scales because liquid motion within the foam has been modeled using the porous media approximation, which assumes immobile bubbles and neglects the kinematic coupling between the solution and the bubbles. This oversight explains the disparity between $\pi_{eff}$ and equilibrium osmotic pressure.
These findings represent significant advancements not only in understanding diffusion phenomena and rheology in foams but also in the behavior of deformable soft particle packing systems (soft jammed systems), such as blood flow in vessels, emulsions, and biological tissues.

Finally, this measurement not only represents a significant advancement in science but also facilitates a rapid assessment of the absorbing capacity of foams. By multiplying both sides of Eq.\ref{eq.Pi} by the base area $S = WD$, we derive the expression $S \pi_{eff} = M_c g$, where $M_c = S H_c \phi_0$ denotes the maximum weight within the foam. Rearranging the equation leads to $\pi_{eff} = M_c g/S$, indicating that $\pi_{eff}$ reflects the maximum mass per unit cross-sectional area over a short time scale. The foam located below the boundary in Fig.\ref{diagram} is capable of absorbing the solution until its mass reaches $M_c$ within this brief time frame. In other terms, the absorbing mass can be estimated as $M_c - M_0$, where $M_0$ signifies the initial mass of the solution within the foam. Therefore, accurately measuring $M_c$ and determining the design parameters for $M_0$ are crucial for optimizing foam products, taking into consideration factors such as foam size, internal bubble size, and liquid fraction.

\section{Summary}
Foams are utilized across various industries, including food production, cleaning agents, cosmetics, construction, and fire extinguishing systems. One of the key physical properties of foams is their absorption capacity, which proves useful in applications such as oil recovery. It has been traditionally assumed that the absorption behavior is governed by the equilibrium osmotic pressure of the foam. Studies on emulsions, systems analogous to foams, have explored equilibrium osmotic pressure using semipermeable membranes~\cite{Cantat2013, Princen1986, Princen1987}, leading to a scaling law that relates this pressure to the liquid fraction, surface tension, and bubble size. However, empirical evidence indicates that the absorption limit is significantly lower than predictions based on equilibrium osmotic pressure. Despite this discrepancy, few experiments have addressed the issue, and the underlying mechanism remains unclear.

In this study, we quantitatively measured the effective osmotic pressure $\pi_{eff}$ on short timescales using three different surfactants. Results reveal that the effective osmotic pressure is approximately two orders of magnitude lower than the equilibrium osmotic pressure. Additionally, observations of the foam's internal structure indicate a correlation between bubble rearrangement and the presence or absence of drainage. Vertically, the solution within the foam flows downward under gravity, causing internal bubbles to shift accordingly. On short timescales, bubbles are subjected to high pressure, which generates a repulsive force that temporarily prevents solution drainage. However, when the internal pressure exceeds the yield stress, bubble rearrangement increases the pressure, permitting drainage. Over longer timescales, as the solution flow diminishes, the foam's internal structure approaches equilibrium, although the jammed state leads to an extremely slow relaxation process.

In foam drainage dynamics, liquid flow and bubble movement are kinetically coupled, creating a transient, strongly non-equilibrium heterogeneous state. This non-equilibrium state is prolonged due to the slow relaxation of the jammed bubbles, resulting in the effective osmotic pressure being determined by the yield stress rather than by equilibrium osmotic pressure. This coupling between liquid and bubble dynamics not only governs drainage and absorption but is also observed in other processes, such as spreading~\cite{Endo2023}. These effects are likely to occur in other soft jammed systems composed of deformable particles, such as emulsions, cells, and biological tissues. The findings of this study offer significant insights into the dynamics of soft jammed systems, particularly regarding the kinetic coupling between dispersed media and dispersed phases.

\section*{Acknowledgements}
R. K. was supported by JSPS KAKENHI Grant Number 20H01874.

\section*{AUTHORS CONTRIBUTIONS}
R.~K. conceived the project. A.~K. performed the experiments and analyzed the data. 
R.~K. wrote the manuscript.

\section*{COMPETING INTERESTS STATEMENT}
The authors declare that they have no competing interests. 

\section*{CORRESPONDENCE}
Correspondence and requests for materials should be addressed to R.~K. (kurita@tmu.ac.jp).

\section*{Availability of Data and Materials}
All data generated or analyzed during this study are included in this published article.


\begin{thebibliography}{10}
\expandafter\ifx\csname url\endcsname\relax
  \def\url#1{\texttt{#1}}\fi
\expandafter\ifx\csname urlprefix\endcsname\relax\def\urlprefix{URL }\fi
\providecommand{\bibinfo}[2]{#2}
\providecommand{\eprint}[2][]{\url{#2}}

\bibitem{Weaire2001}
\bibinfo{author}{Weaire, D.~L.} \& \bibinfo{author}{Hutzler, S.}
\newblock \emph{\bibinfo{title}{The physics of foams}}
  (\bibinfo{publisher}{Oxford University Press}, \bibinfo{year}{2001}).

\bibitem{Cantat2013}
\bibinfo{author}{Cantat, I.} \emph{et~al.}
\newblock \emph{\bibinfo{title}{Foams: structure and dynamics}}
  (\bibinfo{publisher}{OUP Oxford}, \bibinfo{year}{2013}).

\bibitem{Princen1986}
\bibinfo{author}{Princen, H.~M.}
\newblock \bibinfo{title}{Osmotic pressure of foams and highly concentrated
  emulsions. 1. theoretical considerations}.
\newblock \emph{\bibinfo{journal}{Langmuir}} \textbf{\bibinfo{volume}{2}},
  \bibinfo{pages}{519--524} (\bibinfo{year}{1986}).

\bibitem{Princen1987}
\bibinfo{author}{Princen, H.~M.} \& \bibinfo{author}{Kiss, A.~D.}
\newblock \bibinfo{title}{Osmotic pressure of foams and highly concentrated
  emulsions. 2. determination from the variation in volume fraction with height
  in an equilibrated column}.
\newblock \emph{\bibinfo{journal}{Langmuir}} \textbf{\bibinfo{volume}{3}},
  \bibinfo{pages}{36--41} (\bibinfo{year}{1987}).

\bibitem{Mason1997}
\bibinfo{author}{Mason, T.~G.} \emph{et~al.}
\newblock \bibinfo{title}{Osmotic pressure and viscoelastic shear moduli of
  concentrated emulsions}.
\newblock \emph{\bibinfo{journal}{Phys. Rev. E}} \textbf{\bibinfo{volume}{56}},
  \bibinfo{pages}{3150--3166} (\bibinfo{year}{1997}).

\bibitem{Hohler2008}
\bibinfo{author}{H\"{o}hler, R.}, \bibinfo{author}{Sang, Y. Y.~C.},
  \bibinfo{author}{Lorenceau, E.} \& \bibinfo{author}{Cohen-Addad, S.}
\newblock \bibinfo{title}{Osmotic pressure and structures of monodisperse
  ordered foam}.
\newblock \emph{\bibinfo{journal}{Langmuir}} \textbf{\bibinfo{volume}{24}},
  \bibinfo{pages}{418--425} (\bibinfo{year}{2008}).

\bibitem{Yanagisawa2021a}
\bibinfo{author}{Yanagisawa, N.} \& \bibinfo{author}{Kurita, R.}
\newblock \bibinfo{title}{Size distribution dependence of collective relaxation
  dynamics in a two-dimensional wet foam}.
\newblock \emph{\bibinfo{journal}{Scientific reports}}
  \textbf{\bibinfo{volume}{11}}, \bibinfo{pages}{2786} (\bibinfo{year}{2021}).

\bibitem{Yanagisawa2023}
\bibinfo{author}{Yanagisawa, N.} \& \bibinfo{author}{Kurita, R.}
\newblock \bibinfo{title}{Cross over to collective rearrangements near the
  dry-wet transition in foams}.
\newblock \emph{\bibinfo{journal}{submitted}} \textbf{\bibinfo{volume}{13}},
  \bibinfo{pages}{4939} (\bibinfo{year}{2023}).

\bibitem{Keyvan2013}
\bibinfo{author}{Piroird, K.} \& \bibinfo{author}{Lorenceau, E.}
\newblock \bibinfo{title}{Capillary flow of oil in a single foam microchannel}.
\newblock \emph{\bibinfo{journal}{Phys. Rev. Lett.}}
  \textbf{\bibinfo{volume}{111}}, \bibinfo{pages}{234503}
  (\bibinfo{year}{2013}).

\bibitem{Roveillo2020}
\bibinfo{author}{Roveillo, Q.} \emph{et~al.}
\newblock \bibinfo{title}{Trapping of swimming microalgae in foam}.
\newblock \emph{\bibinfo{journal}{The Royal Society Interface}}
  \textbf{\bibinfo{volume}{17}} (\bibinfo{year}{2020}).

\bibitem{Tani2022}
\bibinfo{author}{Tani, M.} \& \bibinfo{author}{Kurita, R.}
\newblock \bibinfo{title}{Pinch-off from a foam droplet in a hele-shaw cell}.
\newblock \emph{\bibinfo{journal}{Soft Matter}} \textbf{\bibinfo{volume}{18}},
  \bibinfo{pages}{2137--2142} (\bibinfo{year}{2022}).

\bibitem{Danov2014}
\bibinfo{author}{Danov, K.~D.}, \bibinfo{author}{Kralchevsky, P.~A.} \&
  \bibinfo{author}{Ananthapadmanabhan, K.~P.}
\newblock \bibinfo{title}{Micelle--monomer equilibria in solutions of ionic
  surfactants and in ionic--nonionic mixtures: A generalized phase separation
  model}.
\newblock \emph{\bibinfo{journal}{Adv. Colloid Interface Sci.}}
  \textbf{\bibinfo{volume}{206}}, \bibinfo{pages}{17--45}
  (\bibinfo{year}{2014}).

\bibitem{Castro1998}
\bibinfo{author}{de~Castro, F. H.-B.}, \bibinfo{author}{G{\'a}lvez-Borrego, A.}
  \& \bibinfo{author}{de~Hoces, M.~C.}
\newblock \bibinfo{title}{Surface tension of aqueous solutions of sodium
  dodecyl sulfate from 20 $\,^{\circ}$c to 50 $\,^{\circ}$c and ph between 4
  and 12}.
\newblock \emph{\bibinfo{journal}{J. Chem. Eng. Data}}
  \textbf{\bibinfo{volume}{43}}, \bibinfo{pages}{717--718}
  (\bibinfo{year}{1998}).

\bibitem{Koehler1998}
\bibinfo{author}{Koehler, S.~A.}, \bibinfo{author}{Stone, H.~A.},
  \bibinfo{author}{Brenner, M.~P.} \& \bibinfo{author}{Eggers, J.}
\newblock \bibinfo{title}{Dynamics of foam drainage}.
\newblock \emph{\bibinfo{journal}{Phys. Rev. E}} \textbf{\bibinfo{volume}{58}},
  \bibinfo{pages}{2097--2106} (\bibinfo{year}{1998}).

\bibitem{deGennes2013}
\bibinfo{author}{De~Gennes, P.-G.}, \bibinfo{author}{Brochard-Wyart, F.} \&
  \bibinfo{author}{Qu{\'e}r{\'e}, D.}
\newblock \emph{\bibinfo{title}{Capillarity and wetting phenomena: drops,
  bubbles, pearls, waves}} (\bibinfo{publisher}{Springer Science \& Business
  Media}, \bibinfo{year}{2013}).

\bibitem{Leonard1965}
\bibinfo{author}{Leonard, R.~A.} \& \bibinfo{author}{Lemlich, R.}
\newblock \bibinfo{title}{A study of interstitial liquid flow in foam. part i.
  theoretical model and application to foam fractionation}.
\newblock \emph{\bibinfo{journal}{AIChE J.}} \textbf{\bibinfo{volume}{11}},
  \bibinfo{pages}{18} (\bibinfo{year}{1965}).

\bibitem{Lorenceau2009}
\bibinfo{author}{Lorenceau, E.}, \bibinfo{author}{Louvet, N.},
  \bibinfo{author}{Rouyer, F.} \& \bibinfo{author}{Pitois, O.}
\newblock \bibinfo{title}{Permeability of aqueous foams}.
\newblock \emph{\bibinfo{journal}{Eur. Phys. J. E}}
  \textbf{\bibinfo{volume}{28}}, \bibinfo{pages}{293--304}
  (\bibinfo{year}{2009}).

\bibitem{Behera2014}
\bibinfo{author}{Behera, M.~R.}, \bibinfo{author}{Varade, S.~R.},
  \bibinfo{author}{Ghosh, P.}, \bibinfo{author}{Paul, P.} \&
  \bibinfo{author}{Negi, A.~S.}
\newblock \bibinfo{title}{Foaming in micellar solutions: Effects of surfactant,
  salt, and oil concentrations}.
\newblock \emph{\bibinfo{journal}{Ins. Eng. Chem. Res.}}
  \textbf{\bibinfo{volume}{48}}, \bibinfo{pages}{18497--18507}
  (\bibinfo{year}{2014}).

\bibitem{Jiang2020}
\bibinfo{author}{Jiang, N.} \emph{et~al.}
\newblock \bibinfo{title}{Role of salts in performance of foam stabilized with
  sodium dodecyl sulfate}.
\newblock \emph{\bibinfo{journal}{Chem. Eng. Sci.}}
  \textbf{\bibinfo{volume}{216}}, \bibinfo{pages}{115474}
  (\bibinfo{year}{2020}).

\bibitem{Weaire2008}
\bibinfo{author}{Weaire, D.}
\newblock \bibinfo{title}{The rheology of foam}.
\newblock \emph{\bibinfo{journal}{Current Opinion in Colloid and Interface
  Science}} \textbf{\bibinfo{volume}{13}}, \bibinfo{pages}{171--176}
  (\bibinfo{year}{2008}).

\bibitem{Endo2023}
\bibinfo{author}{Endo, M.}, \bibinfo{author}{Tani, M.} \&
  \bibinfo{author}{Kurita, R.}
\newblock \bibinfo{title}{Scraping of foam on a substrate}.
\newblock \emph{\bibinfo{journal}{J. Colloid and Interface Sci.}}
  \textbf{\bibinfo{volume}{650}}, \bibinfo{pages}{1612--1618}
  (\bibinfo{year}{2023}).

\end{thebibliography}

\end{document}